\documentclass[pra,twocolumn,showpacs,superscriptaddress,groupedaddress]{revtex4}  
\usepackage{graphicx}
\usepackage{epstopdf}
\usepackage{amsmath,amsthm,amssymb} 
\usepackage{color}
\usepackage{amsfonts}
\usepackage{subfigure}
\usepackage{bm}

\begin{document}

\widetext


\title{Four-level N-scheme crossover resonances in Rb saturation spectroscopy in magnetic fields}


\author{S. Scotto}
\affiliation{Laboratoire National des Champs MagnŽtiques Intenses, (UPR 3228, CNRS-UPS-UJF-INSA),
143 Avenue de Rangueil,  31400 Toulouse, France}
\affiliation{Dipartimento di Fisica ``E. Fermi'', Universit\`a di Pisa, Largo Bruno Pontecorvo 3, 56127 Pisa, Italy}

\author{D. Ciampini}
\affiliation{Dipartimento di Fisica ``E. Fermi'', Universit\`a di Pisa, Largo Bruno Pontecorvo 3, 56127 Pisa, Italy}
\affiliation{INO-CNR, Via G. Moruzzi 1, 56124 Pisa, Italy}
\affiliation{CNISM UdR Dipartimento di Fisica ``E. Fermi'', Universit\`a di Pisa, Largo B. Pontecorvo 3, 56127 Pisa, Italy}

\author{C. Rizzo}
\affiliation{Laboratoire National des Champs MagnŽtiques Intenses, (UPR 3228, CNRS-UPS-UJF-INSA),
143 Avenue de Rangueil,  31400 Toulouse, France}

\author{E. Arimondo}
\affiliation{Laboratoire National des Champs MagnŽtiques Intenses, (UPR 3228, CNRS-UPS-UJF-INSA),
143 Avenue de Rangueil,  31400 Toulouse, France}
\affiliation{Dipartimento di Fisica ``E. Fermi'', Universit\`a di Pisa, Largo Bruno Pontecorvo 3, 56127 Pisa, Italy}\affiliation{INO-CNR, Via G. Moruzzi 1, 56124 Pisa, Italy}
\affiliation{CNISM UdR Dipartimento di Fisica ``E. Fermi'', Universit\`a di Pisa, Largo B. Pontecorvo 3, 56127 Pisa, Italy}
\date{\today}
   \begin{abstract}
   We perform saturated absorption spectroscopy on the D$_2$ line for room temperature rubidium atoms  immersed in magnetic fields within the 0.05-0.13 T range. At those medium-high field values the hyperfine structure in the excited state is broken by the Zeeman effect, while in the ground state hyperfine structure and Zeeman shifts are comparable. The observed spectra are composed by a large number of absorption lines. We identify them as saturated absorptions on two-level systems, on three-level systems in a V configuration and on four-level systems in a N or double-N configuration where two optical transitions not sharing a common level are coupled by spontaneous emission decays.  We analyze the intensity of all those transitions within a unified simple theoretical model. We concentrate our attention on the double-N crossovers signals whose intensity is very large because of the symmetry in the branching ratios of the four levels. We point out that these structures, present in all alkali atoms at medium-high magnetic fields, have interesting properties for electromagnetically induced transparency and slow light applications.  
     \end{abstract}

\maketitle

\section{Introduction}
The laser action on atoms modifies the level occupation, and in addition may produce coherences associated to the quantum mechanical superposition of atomic states. The development of laser spectroscopy tools and the availability of samples with new features has allowed the exploration of a larger set of coherent and  incoherent processes. Moving from two-level systems to three-level ones, phenomena as coherent population trapping, electromagnetically induced transparency, slow light were discovered and largely exploited~\cite{Fleischhauer}. This trend was confirmed by the search for the new properties of four-level schemes, forming the so-called N-configuration,  with the  first proposition by Harris~\cite{Harris1998}, followed by the experimental investigation by several authors, as~\cite{YanZhu2001,BrajeHarris2004,Kang2004,ChenYu2005},  with lineshape contributions more recently explored by~\cite{Abi-Salloum2010}.\\
\indent Here we explore the high-resolution absorption spectra of rubidium atoms in medium-high magnetic fields. At low magnetic fields, i.e., in presence of small Zeeman splittings of the atomic levels, the alkali absorption spectra should be treated as a multilevel systems, with a very complex interaction with the laser, typically eliminated by optical pumping into the extreme cycling  transitions. This complexity is washed out at the 0.05-0.13 T magnetic fields of our investigation  where the Zeeman effect breaks all degeneracies and a very large number of isolated and open two-level systems  describe the laser-atom interaction.  At those medium-high magnetic fields the hyperfine splitting is broken by the magnetic field in the excited state, but not in the ground state, and the usual optical dipole selection rules are broken by the magnetic field mixing. Therefore  the spectra are characterised by several well isolated three-level structures. In addition some peculiar, and very strong, features associated to different four-level N schemes are produced by the laser excitation. Similar strong absorption structures shoould appear in other alkalis spectra at those medium-high fields. \\ 
\indent  Sub-Doppler absorption signals produced by four-level systems, within an N configuration, were studied a long time ago, as soon the sub-Doppler saturated spectroscopy was introduced, see~\cite{Cahuzac1977,Nakayama1980}, with a review in~\cite{Nakayama1997}. Those studies were performed at magnetic fields lower than those of the present work, and  therefore our new features were not observed. Let's point out that in all our four-level schemes spontaneous emission coupling represents an intermediate step within  the N structure. Therefore only a population transfer occurs, without  creation of atomic coherences, unlike the coherent N schemes mentioned above triggered by the~\cite{Harris1998} proposal. The large variety of four-level schemes associated to our intermediate magnetic field regime provide to experimentalists a flexible medium where to implement different coherent processes. At some magic magnetic field values and  by replacing the spontaneous emission with stimulated emission (produced by close frequency additional lasers), our  schemes may be converted into peculiar double-$\Lambda$ structures.\\ 
\indent  Saturated absorption resonances in alkali atom spectra in presence of magnetic fields larger than 0.01 T were examined by several authors, also within the last few years, see for instance~\cite{Tremblay1990,Momeem2007,Skolnik2009,OlsenHapper2011,HakhumyanSarkisyan2012,SargsyanMariottiSarkisyan2014}. Because our study of Zeeman effect in rubidium complements those works, our analysis includes the observation of absorption lines forbidden at zero magnetic field, but allowed  increasing the field owing to the magnetic field induced mixing of the eigenstates. Our attention is focused on the strength of the saturated absorption signals associated to two-level schemes (TLS), V-scheme three-levels (VTL), and  four-level N schemes. Our list does not include $\Lambda$ three-level systems, because they are not present within the explored  laser polarization schemes. \\
\indent The theoretical description of saturation spectroscopy has a long history, with an early work by H\"ansch~\cite{Haensch1977} providing an introductory review. The standard approach is based on the steady state solutions of the density matrix equations for the atomic levels excited by the control and probe lasers, as initiated in~\cite{Holt1972}. On the basis of that formalism Nakayama {\it et al.}~\cite{Nakayama1980,Nakayama1997} derived expressions for the signal intensities. The integration over the Doppler velocity distribution represents a key element for quantitative analysis, as  in~\cite{PappasFeld1980,DoJhe2008,MoonNoh2008,NohJhe2010}. Our Zeeman spectra are composed by a large number of absorption lines, the exact number depending on the laser polarization scheme, and their largest majority is associated to two-level open systems, where the optical pumping into a dark state determines the signal intensity.  The contribution of optical pumping into dark states  was examined in~\cite{PappasFeld1980,SmithHughes2004}. We compare the relative amplitude of the absorption signals  by using simple and  universal formulas without performing a Doppler integration as required for precise amplitude determinations.   The role of the transient optical pumping  on the amplitude of the saturated absorption signal discussed in~\cite{PappasFeld1980,SydorykEkers2008} is important for some experimental observations.\\ 
\indent In Section II we rewrite the steady-state density matrix solution in a form  appropriate to analyze open TLS's. The following Section III derives from that solution the intensities of the  VTL and N absorption features. After a brief Section presenting our apparatus, Section V discusses typical spectra, and points out some observed forbidden lines. Section VI analyzes the intensity of all the measured absorption signals. A perspective and conclusion Section completes our work.\\
\begin{figure}[!b]
\centering
\includegraphics[width=5 cm]{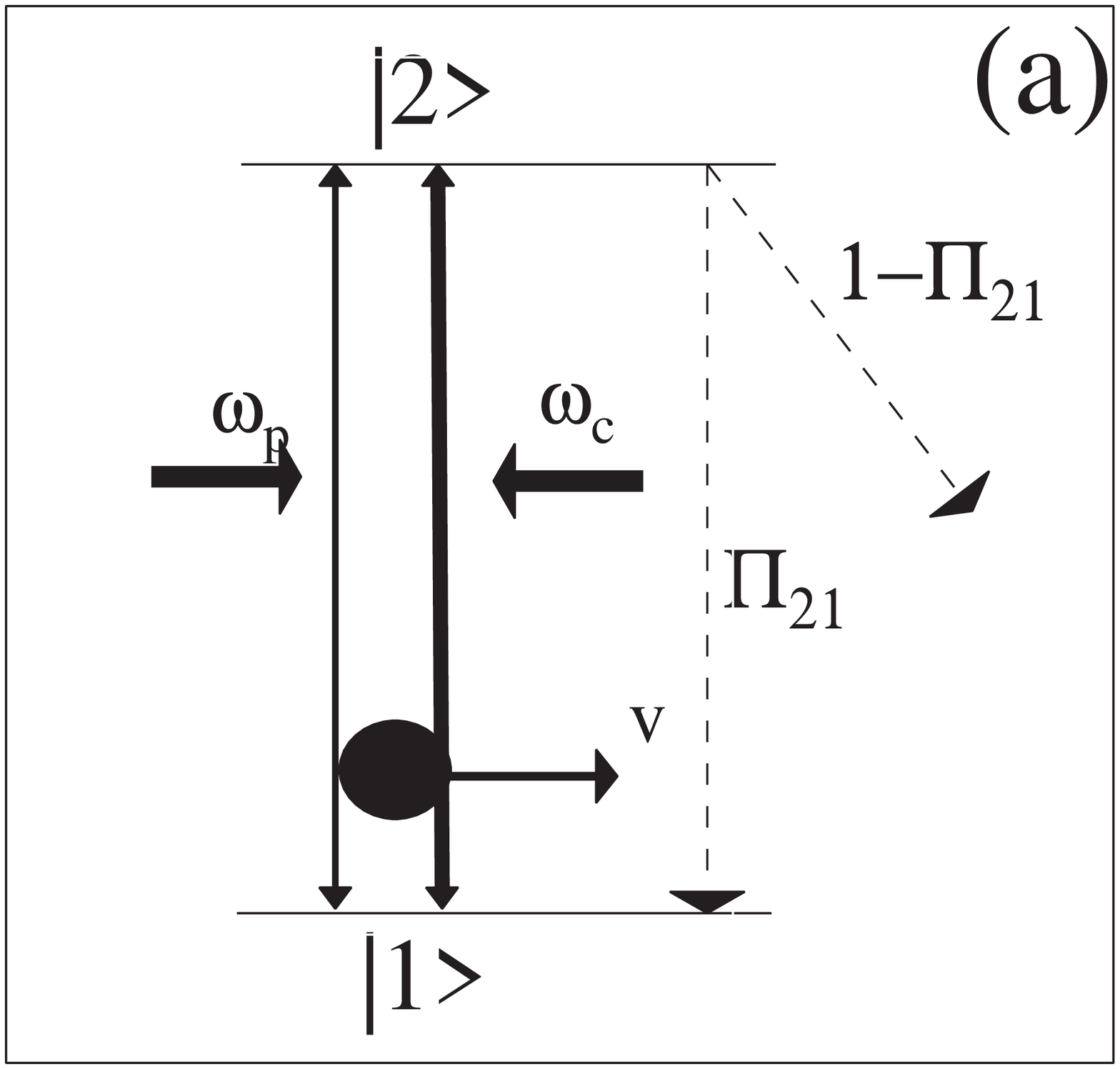}
\includegraphics[width=7 cm]{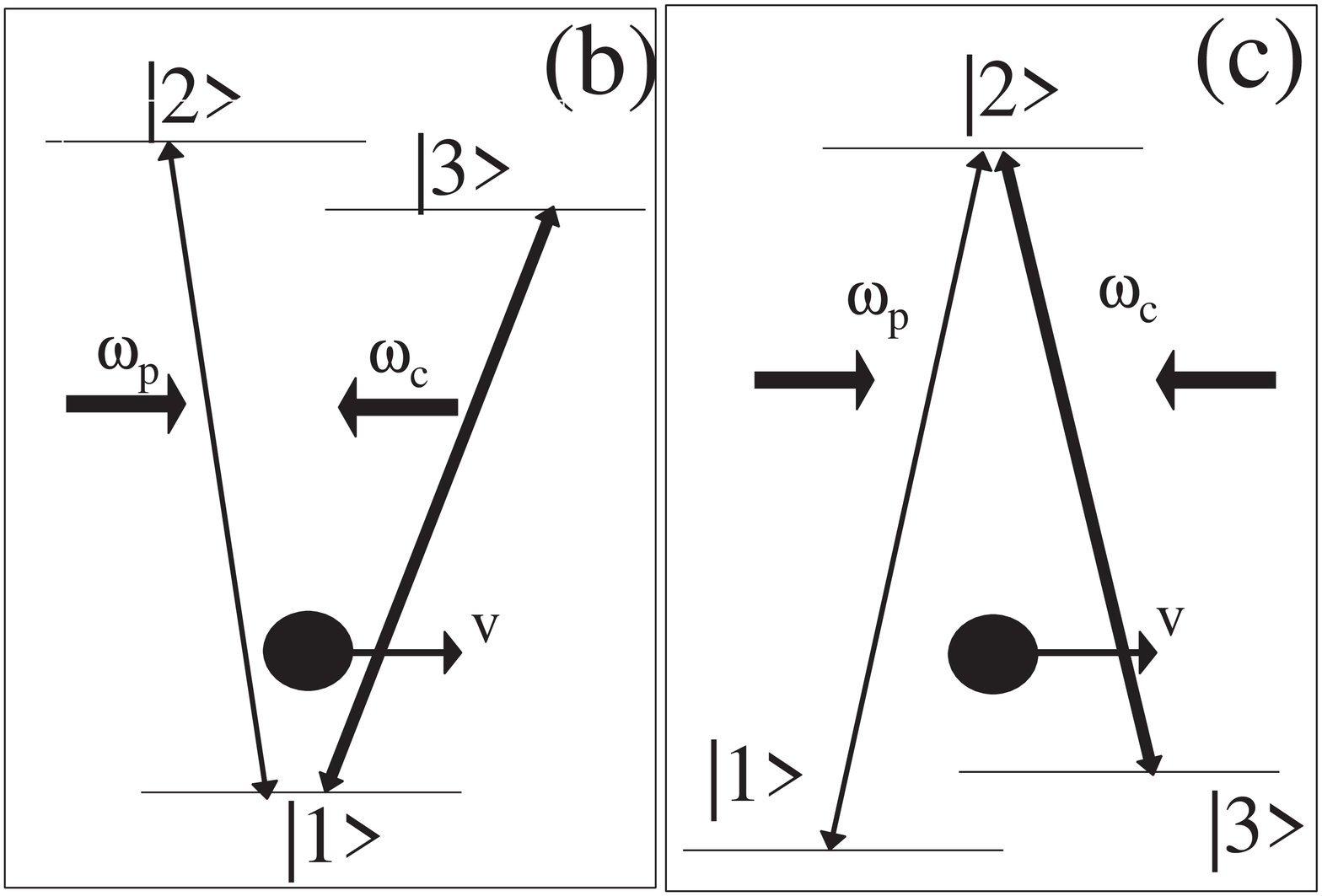}
\includegraphics[width=7 cm]{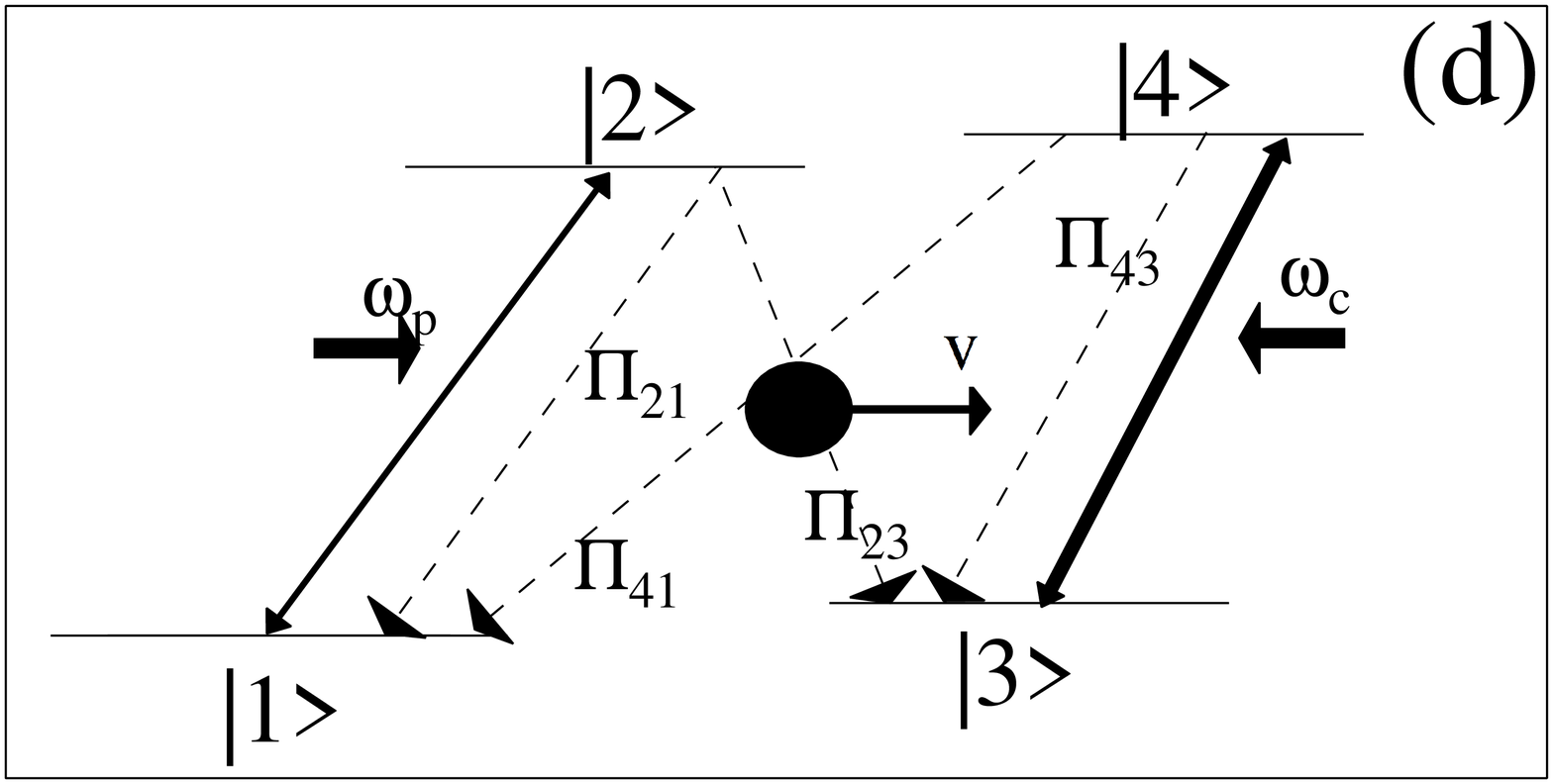}
\caption{Level schemes for the sub-Doppler saturation spectroscopy:  two-level  scheme in a), three-level V scheme in b) and $\Lambda$ scheme in c), and N four-level scheme in d). Strong control laser at angular frequency $\omega_c$ and weak probe laser at angular frequency $\omega_p$ traveling in opposite directions excite the atomic levels.  In a) and d) the spontaneous emission coupling between levels $i \to j$ is  described by the branching ratios  $\Pi_{ij}$.}
\label{Energylevels}
\end{figure}
 \section{Two-level open system}  
 \indent In order to compare the saturation absorption signals associated of a variety of optical transition, for each  given $|1\rangle \to |2\rangle$ TLS we introduce a theoretical relative line strength $f_{1-2}$  defined as the square of the dipole moment $D$ matrix element between initial and final states, normalized to the optical dipole matrix element for the strong cycling Zeeman transition of the $5^2$S$_{1/2} \to$5$^2$P$_{3/2}$ D$_2$ line.  Thus for a $^{87}$Rb transition we write
 \begin{equation}
 f_{1-2}=\frac{|\langle 1|D|2\rangle |^2}{|\langle 5^2\mathrm{S}_{1/2}, F_g=2,m_g=2|D|5^2\mathrm{P}_{3/2}F_e=3,m_e=3\rangle |^2}
 \end{equation}
Therefore $f_{1-2}$ is equal to one for the cycling transition and smaller than one for all the open ones.\\
\indent In  saturation spectroscopy atoms moving in random directions are irradiated by two laser beams, control and probe of frequencies $\omega_c$ and $\omega_p$, with  intensities $I_c$ and $I_p$, respectively. The strong control laser modifies the atomic population, and the weak probe laser monitores those modifications. The saturation spectroscopy eliminates Doppler broadening by using control and probe beams traveling in opposite directions. Because of Doppler shift the atoms having in the laboratory frame an axial velocity $v$ experience the  control and probe laser beams shifted in frequency to $\omega_p-kv$ and  $\omega_c+kv$, with $k$ the laser wavenumber, as schematized in Fig.~\ref{Energylevels} for different atomic level schemes. Only atoms moving in a narrow velocity range along the optical axis interact simultaneously with control and probe lasers.\\
\indent  To analyze our results we solve the Optical Bloch equations in presence of the control laser and use that solution to calculate the absorption of the probe laser. An important ingredient is the openness of the transitions driven by the two lasers. For the simplest case of a two-level transition as in Fig.~\ref{Energylevels}(a) that character is described by the branching ratio $\Pi_{21}$ from the upper $|2\rangle$ level to the lower one, and by the complement $1-\Pi_{21}$ towards other levels.\\

\subsection{Steady state regime}
\indent  Under the control laser excitation,  by eliminating the atomic coherences within the Optical Bloch equations as in~\cite{PappasFeld1980,SydorykEkers2008}, for a given atomic velocity $v$ the $\rho_{11},\rho_{22}$ populations within the $|1\rangle$ and $|2\rangle$ levels, respectively,  of Fig.~\ref{Energylevels}(a)  are given by 
\begin{eqnarray}
\frac{\partial}{\partial t}\rho_{11}&=& \Gamma_c\left(\rho_{22}-\rho_{11}\right)+\Gamma\Pi_{21} \rho_{22} -\gamma\left(\rho_{11}-\rho_{11}^0\right) \nonumber \\
\frac{\partial}{\partial t}\rho_{22}&=&-\Gamma_c\left(\rho_{22}-\rho_{11}\right) -\Gamma\rho_{22} 
\label{TLSequations}
\end{eqnarray}
with  $\Gamma$ the excited state spontaneous decay rate, and $\gamma$ the rate at which atoms are produced in the ground state $|1\rangle$. The steady state occupation $\rho_{11}^0(v)$ is given by 
\begin{equation}
 \rho_{11}^0(v)= W(v)=\frac{1}{\sqrt{\pi}u}e^{-\left(\frac{kv}{\Delta_D}\right)^2}.
\end{equation}
$W(v)$ is the Gaussian atomic velocity distribution, with $\Delta_D=ku$ the Doppler width, $u= \left(2k_BT/M\right)^{1/2}$ the most probable velocity, $T$ the gas temperature, $k_B$ the Boltzmann constant,  and $M$ the atomic mass. $\Gamma_c$ is the control laser pumping rate~\cite{SydorykEkers2008}
\begin{equation}
\Gamma_c=(\Omega^c_{1-2})^2\frac{\Gamma}{\Gamma^2+4\Delta_c^2}=f_{1-2}\frac{I_c}{2I_{sat}}\frac{\Gamma}{1+4\Delta_c^2/\Gamma^2},
\end{equation}
where $\Omega^c_{1-2}$ is the laser Rabi frequency and $\Delta_c$ the laser detuning for the driven transition. The control pumping rate has been rewritten on the basis of the  $f_{1-2}$ line strength, and of the  $I_{sat}$ saturation intensity for the B=0 cycling transition reported in \cite{Steck}.\\
\indent The $I_p$ laser probes the $|1\rangle-|2\rangle$ population difference   perturbed by the control laser. The steady state population difference derived from Eqs.~\eqref{TLSequations} determines the probe laser absorption coefficient and the amplitude  of the TLS saturated absorption signal. For a weak probe laser, the  absorption is proportional  to the $I_p$ probe intensity. For coupling/probe resonant frequencies $\omega_c=\omega_p=\omega_{21}$ and a given $I_p$ laser intensity we characterize the absorption amplitude through the following dimensionless $f^{TLS}$ parameter:
\begin{equation}
f^{TLS} = - f_{1-2}\frac{I_c}{I_c+I_{op}},
\label{TLSstrength}
\end{equation}
depending on the  $f_{1-2}$ strength of the absorbing transition. The $I_{op}$ optical pumping saturation intensity in~\cite{PappasFeld1980}, or depletion saturation intensity in~\cite{SydorykEkers2008}, is given by
\begin{equation}
I_{op}=\frac{I_{sat}}{f_{1-2}}\frac{1+\frac{\gamma}{\Gamma}}{1+\frac{\Gamma(1-\Pi_{21})}{2\gamma}}.
\end{equation}
For our present target of comparing different TLS strengths,  the $f^{TLS}$ parameter does not include the  $W(0)=(\sqrt{\pi}u)^{-1}$ fraction of absorbing atoms at zero velocity. The negative sign within Eq.~\eqref{TLSstrength}  implies that the atomic absorption is decreased by the control laser saturation. Most experimental observations are performed at  $I_c \gg I_{op}$ and therefore $|f^{TLS}|\approx f_{1-2}$.\\
\subsection{Transient regime}
Within the optical pumping process a key role is played by the $\tau_{op}$ optical pumping time~\cite{SydorykEkers2008} 
\begin{equation}
\tau_{op}=\frac{1+\Gamma_c/\Gamma}{\Gamma_c\left(1-\Pi_{21}\right)}.
\end{equation} 
If $\tau_{op}$ is comparable or larger than $1/\gamma$, the above steady state approach is not valid. In~\cite{SydorykEkers2008}  the saturated absorption signal is obtained by  imposing $\gamma=0$ within Eqs.~\eqref{TLSequations}, and integrating from 0 to $1/\gamma$ their time dependent solution. The final result for the $f$ parameter within this transient regime is
\begin{equation}
f^{TLS}_{tran} = -f_{1-2}\left[1-\frac{I_{sat}}{f_{1-2}I_c\left(1-\Pi_{21}\right)}\left(1-e^{-\tau_{op}\gamma}\right)\right],
\label{TLStransient}
\end{equation}
Notice that both the optical pumping saturation intensity and the pumping time play a key role for transitions with $\Pi_{21}$ close to one. 
\section{Crossover saturation signals}
Crossover (CO) saturation spectroscopy has been investigated by a large number of authors in a variety of level schemes. Our attention is concentrated on the  VTL and N configurations represented in Figs.~\ref{Energylevels}b) and d).  For the VTL configuration the control and probe lasers act on two transitions $|1\rangle \to |2\rangle$ and $|1\rangle \to |4\rangle$, respectively, sharing the lower state $|1\rangle$. For the N four-level scheme control and probe act on two separate transitions  $|1\rangle \to |2\rangle$ and $|3\rangle \to |4\rangle$ and a coupling between the laser actions is produced by spontaneous emissions processes $|2\rangle \to |3\rangle$ and/or $|4\rangle \to |3\rangle$ with branching ratios $\Pi_{23}$ and $\Pi_{41}$. Both processes were identified for the first time in refs.\cite{Cahuzac1977,Nakayama1980}. CO signals produced by $\Lambda$ saturation scheme of Fig.~\ref{Energylevels} c) are not observed in our $\pi\pi$ and $\sigma\sigma$ polarizations  because for our magnetic fields the ground state Zeeman splittings are larger than the Doppler width and no atoms are available for $\Lambda$ schemes.\\  
\subsection{V-scheme crossover}
 The   $f^{VTL}$ signal  amplitude of the VTL signal is derived by including an additional upper level to the TLS analysis of  previous Section. Extending that analysis to probe on the $|1\rangle \to |2\rangle$ transition and control on the   $|1\rangle \to |3\rangle$, we obtain for   the resonant steady state $f^{VTL}$ amplitude 
 \begin{equation} 
f^{VTL} = -\frac{W(v)}{W(0)}f_{12}\frac{\left(1+\frac{\Gamma}{\gamma}(1-\Pi_{31})\right)f_{13}I_c}{\left(1+\frac{\gamma}{\Gamma}\right)I_{sat}+\left(1+\frac{\Gamma}{\gamma}\frac{1-\Pi_{31}}{2}\right)f_{13}I_c}.
\label{VTLamplitude}
\end{equation}
with the Gaussian $W(v)$ probability calculated at the  atomic velocity $v$ given by  
\begin{equation}
v= \frac{1}{2k}\left[\left(\omega_{31}-\omega_{c}\right)-\left(\omega_{21}-\omega_p\right)\right],
\label{veloccond}
\end{equation}
and resonance condition
\begin{equation}
\omega_c+\omega_p=\omega_{21}+\omega_{31}.
\label{resonance}
\end{equation}
The  $W(v)/W(0)$ fraction in Eq.~\eqref{VTLamplitude} describes the strength reduction  owing to the reduced number of absorbing atoms at velocity $v$. It is included in order to compare the $f^{TLS}$ parameters to  the corresponding ones for the multilevel systems. The minus sign  indicates that saturated VTL produces a decreased absorption. Owing to the investigated regime  $I_p \ll I_c$ our analysis neglects the contribution of the coherence between the upper levels~\cite{Saprykin2012}, because the laser intensity $I_p$ is very small as compared to $I_c$.    As pointed in~\cite{Nakayama1980,Nakayama1997}, within a three-level system the role of control and probe lasers may be reversed. Therefore two different velocity classes, given by Eq.~\eqref{veloccond} and its opposite, determine the total VTL  signal amplitude.  \\

\begin{figure*}[!ht]
\centering
\includegraphics[width=18 cm]{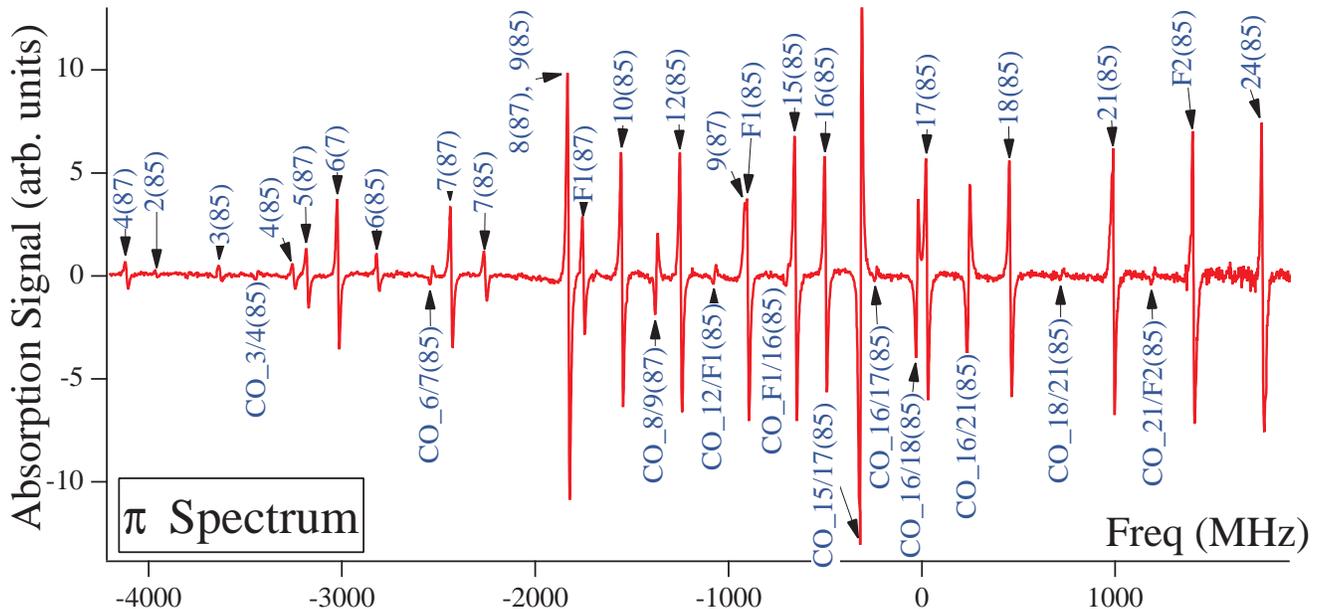}
\caption{Derivative of the rubidium sub-Doppler absorption spectrum vs the laser frequency for an applied $B=0.126$ T field and $\pi$ polarization of both control/probe laser beams. The spectrum covers a 6 GHz wide spectral range, limited by laser instabilities on the high frequency side. The zero frequency is the centre of gravity of the $B=0$ $^{85}$Rb spectra. The transitions are numbered following their increasing frequency position at $B=0$, with the isotope number in parenthesis at the end. For the zero-field forbidden transitions, F1 and F2, of $^{85}$Rb and the F1 one of $^{87}$Rb appearing within the spectrum, the quantum numbers are reported within the text. All the appearing CO's are produced by N schemes with the an increased absorption, and the -316 MHz strongest one is produced by a double-N level scheme.}
\label{Spectrum}
\end{figure*}

 \subsection{N and double-N crossovers} 
In a single N-scheme  the probe laser acts on the $|1\rangle \to |2\rangle$ transition and the control on the   $|3\rangle \to |4\rangle$  one with one-way coupling $\Pi_{41}\ne0$, i.e., $\Pi_{23}=0$, see Fig.~\ref{Energylevels}(d).  The resonant $f^{N}$ crossover amplitude  is given by  
\begin{equation}
f^{N} = \frac{W(v)}{W(0)}f_{12}\frac{\Pi_{41}f_{34}I_c}{\frac{\gamma}{\Gamma}\left(1+\frac{\gamma}{\Gamma}\right)I_{sat}+\left(\frac{1-\Pi_{43}}{2}+\frac{\gamma}{\Gamma}\right)f_{34}I_c}
\label{Namplitude}
\end{equation}
$W(v)$ corresponding to the resonant atomic velocity 
\begin{equation}
v= \frac{1}{2k}\left[\left(\omega_{43}-\omega_{c}\right)-\left(\omega_{21}-\omega_p\right)\right],
\label{veloccond2}
\end{equation}
and resonance condition
\begin{equation}
\omega_c+\omega_p=\omega_{21}+\omega_{43}.
\label{resonanceN}
\end{equation}
 Again the  $W(v)/W(0)$ fraction describes the strength reduction  due to the reduced number of absorbing atoms at velocity $v$. The positive sign indicates that the N saturation signal corresponds to an increased absorption. Within the single N scheme the role of the control and probe lasers cannot be reversed and only one velocity class contributes to the signal.\\
\indent The double-N crossovers are associated to a four level structure where  spontaneous emission decays are present in both directions, i.e., both $\Pi_{23}$ and $\Pi_{41}$ are different from zero. Thus the overall level structure is equivalent to two separate N schemes, (1-2-3-4) and (3-4-1-2) where the role of the control and probe lasers is reversed. Two different velocities classes at $v$ given by Eq.~\eqref{veloccond2} and at  its opposite contribute to the absorption signals. The double-N amplitude is obtained by summing up the  $f^N$ amplitudes of each N configuration. The medium-high magnetic field spectrum of rubidium atoms contains some very strong double-N crossovers where a  complete symmetry exists between the levels. The two transitions driven by the control and probe lasers have comparable transition strengths, i.e., $\Pi_{21}\approx \Pi_{43}$, and that applies also to the decay rates, i.e.,  $\Pi_{23}\approx1-\Pi_{21}\approx\Pi_{41}\approx1-\Pi_{43}$.    

\section{Apparatus}
\label{apparatus}
\begin{figure}[!t]
\centering
\includegraphics[width=9 cm]{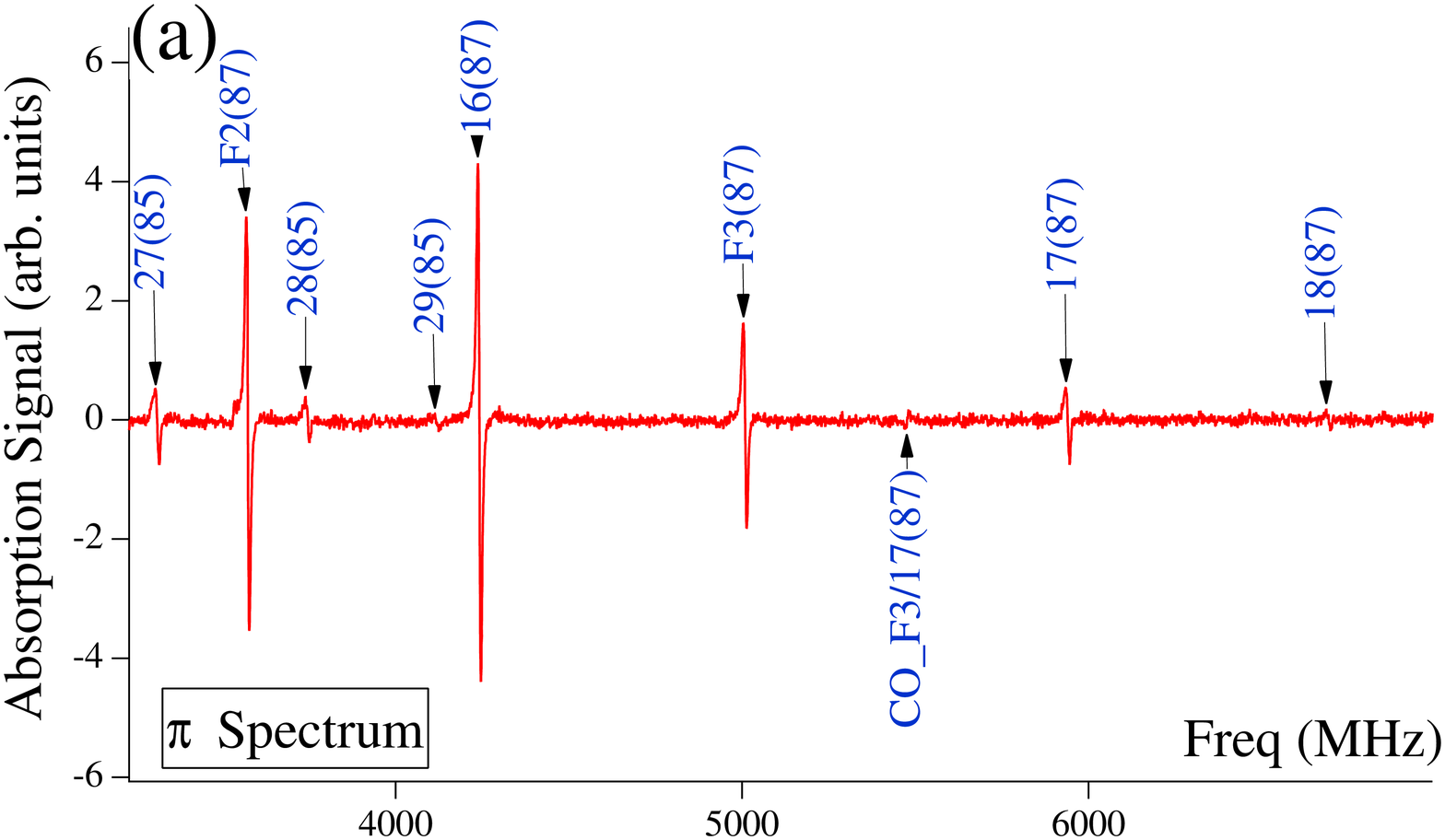}
\includegraphics[width=5 cm]{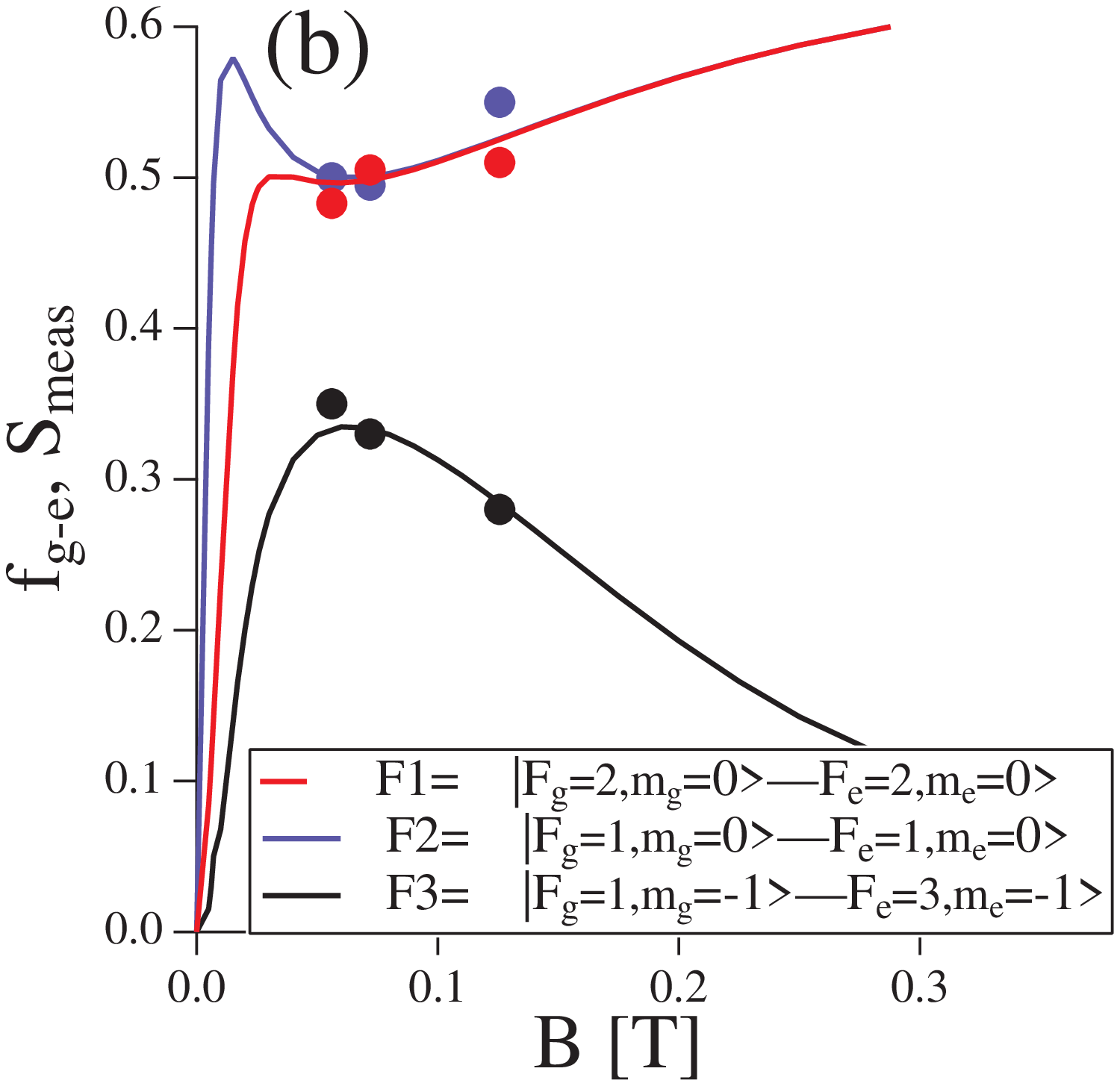}
\caption{(a) Spectra for the same conditions of Fig. 2 with two, F2 and F3, $^{87}$Rb   zero-field forbidden  lines. (b) Line strength $f_{g-e}$ vs magnetic field $B$ for the $^{87}$Rb F1-3 lines with their $B=0$ quantum numbers reported in the inset. The dots denote the $S_{meas}$ experimental observations  normalized to the black line value at 0.072 T. Ten percent indetermination associated to the experimental values.}
\label{Forbidden}
\end{figure}

The experiments are performed with a commercial grating-feedback laser diode (Toptica Photonics, DLX110) operating in Littrow configuration. The 780 nm laser diode, stabilized in temperature, has a typical 1 MHz free-running  linewidth and an output power up to 500 mW, but after optical fibre cleaning only a power less than  5 mW is used for each of the saturated absorption set-ups listed in the following. Part of the laser emission is sent into a homemade FabryÐPerot interferometer (FPI), with 200 MHz free-spectral-range and 250 finesse, whose transmission peaks provide a frequency reference for the measurements of rubidium spectral features. A commercial Wavemeter was used for very precise frequency measurements. \\
\indent The laser power is injected into two saturated absorption spectroscopy set-ups, one containing a rubidium cell placed in the lab background magnetic field, and the second one containing a cell placed within permanent magnets. The rubidium homemade vapor cells contain $^{85}$Rb and $^{87}$Rb in natural abundance. The control beam is frequency modulated by passing through an AOM in double passage and lock-in detection at the 60 KHz modulation frequency is applied to the probe transmitted light. The diameter probe laser beam, around 3 mm leading to  $\gamma=1\times10^5$ s$^{-1}$,  is smaller than the control laser one. \\
\indent A 8-pole Halbach-like~\cite{Halbach1980} magnetic field configuration, similar to those described in  \cite{Cheiney2011,DanieliCasanova2013,HaaHama2014}, provides a uniform field within the laser-atom interaction region. It is  created by assembling on a cylindrical configuration several small NdFeB permanent magnets, with a 1.08 T remnant magnetic field. Magnet assemblies with different radial dimensions are fabricated to produce the magnetic fields explored in this work. Numerical investigations as in \cite{Cheiney2011}  indicate that our assembly  is able to produce fields up to few Tesla with an homogeneity around $10^{-3}$ T within the atomic volume illuminated by the laser, 3 mm in diameter and 20 mm length.  \\
\indent Saturated absorption spectra on the D$_2$ transition are recorded by scanning the laser frequency, and examples are in Figs.~\ref{Spectrum} and \ref{Forbidden}(a). The absorption lines appear with a derivative lineshape because of the lock-in detection.  The FPI output port allow us to monitor the laser power over the frequency scan and to correct the spectrum intensity for variations of the $I_p$ intensity.   For each absorption line we determine the relative amplitude $S_{meas}$  defined as the integrated intensity of the saturated absorption.  Because the absorption derivatives are recorded, see  Figs.~\ref{Spectrum} and \ref{Forbidden}(a), $S_{meas}$ is obtained by fitting the measured signal to the derivative of an absorption Lorentzian lineshape. In the case of strong absorption lines we introduce into $S_{meas}$  a correction for the sample optical tickness. We take the B=0 $^{87}$Rb $|F_g=2,m_g=2\rangle \to |F_g=3,m_g=3\rangle$ saturated signal as a reference signal with  $S_{meas}=1$.   Because the operations of removing the magnet assembly and replacing with a B=0 cell are not systematically performed, the $S_{meas}$ values have an estimated  ten percent indetermination.

\section{Spectra}
For the  investigations reported here with magnetic fields in the range  between 0.05 and 0.13 T, the hyperfine coupled basis $|F_g,m_g\rangle_g$ is perfectly valid for the atomic ground state, but it is not  exact for the excited state, where the magnetic interaction is larger than the hyperfine coupling. However following the seminal work by Tremblay et al.~\cite{Tremblay1990} that explored the rubidium absorption in a similar magnetic field range, we adopt the hyperfine coupled basis also for the excited state and denote the transitions as $|F_g,m_g\rangle \to  |F_e,m_e \rangle$. \\
\indent Saturated absorption spectra   recorded with $\pi$ polarizations of both control and probe laser for an applied magnetic field $B=0.126$ T are shown in Figs.~\ref{Spectrum} and 3(a).  These spectra present a quite large number of lines, and the complexity increases when the $\sigma/\sigma$ polarization  configuration is explored. We have not studied the $\pi/\sigma$ configuration where the spectrum is even more crowded. The spectra contain lines with opposite sign. All lines produced by saturated absorption with control and probe acting on the same two-level system, as in Fig.~\ref{Energylevels}(a), have the same sign. They are numbered in the upper part of the spectra, with the isotope indication in parenthesis at the end. The reported numbers denote their position in the spectra at increasing frequency. Because of the different Zeeman splittings in the ground and excited states, the $B=0$ sequence does not match always the frequency order at the large magnetic fields of the present investigation.\\
 \indent Few two-level lines denoted as  F (forbidden) are present in both spectra. They are forbidden at $B=0$ because of the optical dipole selection rules. At intermediate magnetic field they become allowed by the magnetic field mixing of the hyperfine eigenstates, as recently investigated for rubidium in~\cite{HakhumyanSarkisyan2012} and for caesium in~\cite{SargsyanMariottiSarkisyan2014}.  Within the magnetic field range of the present investigation, the forbidden lines acquire a large probability, as evidenced by the those appearing in the $B=0.126$ T $\pi$  spectra of  Figs.~\ref{Spectrum} and ~\ref{Forbidden}(a). The theoretical and measured intensities of the F1-3 lines of $^{87}$Rb, with their quantum numbers,   are  reported in Fig.~\ref{Forbidden}(b) as function of the magnetic field. The  forbidden lines for the $^{85}$Rb isotope appearing within the spectrum of Fig.~\ref{Spectrum}  are identified as F1(85)$=|3,0\rangle_g \to |3,0\rangle_e$ and F2(85)$=|2,0\rangle_g \to |2,0\rangle_e$.\\
 \begin{figure}[!t]
\includegraphics[width=9 cm]{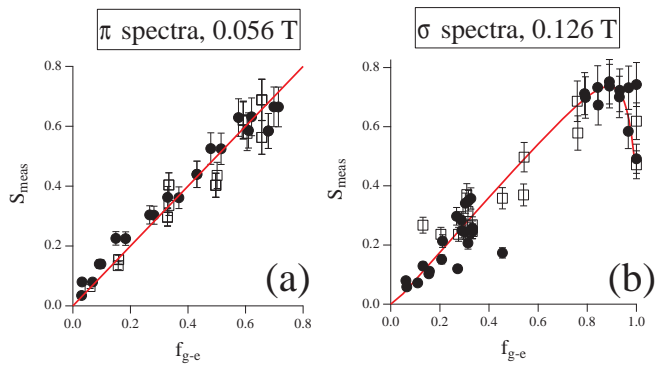}
\caption{Comparison between the $S_{meas}$ intensities of VTL and N  saturation signals (data points) and  the $f^{TLS}$ theoretical predictions (continuous red lines) vs the $f_{g-e}$ of the absorption line. Closed circles for $^{85}$Rb data and open squares for $^{87}$Rb ones. In (a) $\pi/\pi$ polarization signals at 0.056 T, in (b) $\sigma/\sigma$ polarization signals at 0.126 T. Theoretical predictions derived from Eq.~\eqref{TLSstrength} in (a), and from Eq.~\eqref{TLStransient} in (b) with $I_c=2I_{sat}$ and $\gamma=1.10^5$ s$^{-1}$.}
\label{Fig4}
\end{figure}
 \indent  The remaining spectra lines, labeled as COxx in the bottom, are associated to CO level schemes. The CO features are determined by the two optical transitions, and their labels  are inserted into each CO label. The CO laser resonance condition is the median of the participating transitions, as easily verified. That condition is very useful in deriving the quantum numbers of the participating levels. All CO signals of both reported spectra are produced by the four-level configuration of  Fig.~\ref{Energylevels}(c). They correspond to an increased absorption and therefore have a sign opposite to that of the two-level signals. The spectrum of Fig.~\ref{Spectrum} is characterized by a very strong CO at -316 MHz frequency produced by a very symmetric double-N scheme, $\Pi_{21}\approx\Pi_{43}\approx0.6$ and  $\Pi_{23}\approx\Pi_{41}\approx0.3$. \\
 \indent CO's associated to the $V$ scheme of Fig.~\ref{Energylevels}(b) produce absorption signals with the same sign as the two-level ones. However at $B=0.126$  T very few ones were detected   because for most existing V-schemes the frequency  separation between the upper levels is much larger than the $\approx$500 MHz Doppler width, and very few atoms have the resonant velocity required by Eq.~(\ref{veloccond}). V-scheme CO's  were detected in the spectra recorded at lower magnetic fields. For instance, a 0.072 T  wide scanning $\sigma/\sigma$ spectrum contained 82 absorption lines, with  27 N  and 1 VTL CO's.\\ 
 
\begin{figure}[!t]
\centering
\includegraphics[width=9 cm]{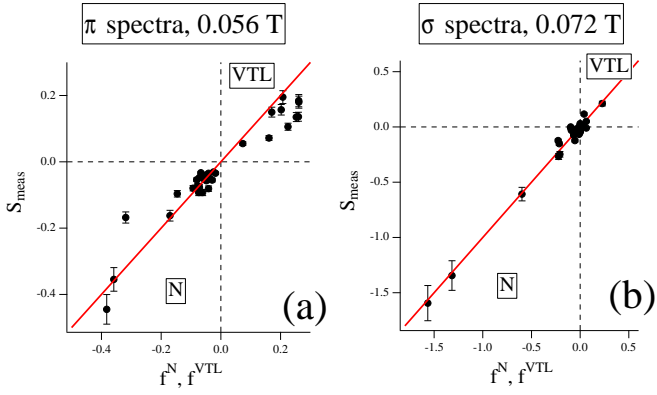}
\caption{$S_{meas}$ intensities of VTL and N  saturation signals (data points) vs  $f^{VTL}$ or $f^{N}$. Positive values correspond to VTL signals, described by Eq.~\eqref{VTLamplitude},  and negative values correspond to N signals, described by  Eq.~\eqref{Namplitude}.  In (a) signals at 0.056 T in a $\pi/\pi$ spectrum, in (b) signals at 0.072 T in a $\sigma/\sigma$ spectrum. The  red lines report theoretical predictions based on the TLS linear dependence of Fig.~\ref{Fig4}(a).}
\label{Fig5}
\end{figure} \section{Line strengths}
 \label{strengths}
 Fig.~\ref{Fig4}(a)  reports the $S_{meas}$ measurements of TLS on a $\pi/\pi$ spectrum at  0.056 T, for a control laser intensity $I_c=2I_{sat}$, equivalent to $I_c \to \infty$ in Eq.~\eqref{TLSstrength}. Therefore  $|f^{TLS}|\approx f_{g-e}$ and the  linear dependence of $S_{meas}$ on $f_{g-e}$ predicted by that equation is confirmed.  Fig.~\ref{Fig4}(b)  reports the TLS measurements on a $\sigma/\sigma$ spectrum at  0.126 T at the same control laser intensity in presence of transitions with a large variety of $f_{g-e}$ values, and several of them close to 1. Notice that for those values the linear dependence prediction is not satisfied. Instead  a dependence of $S_{meas}$ on $f_{g-e}$ with an intermediate  maximum is predicted by the treatments of both refs.~\cite{PappasFeld1980} and \cite{SydorykEkers2008} including the role of the transient regime. Fig.~\ref{Fig4}(b) shows the good theory-experiment agreement using Eq.~\eqref{TLStransient}. Notice that the deviation from the linear regime appears only for $f_{g-e}\ge 0.8$ and therefore the data of  Fig.~\ref{Fig4}(a) are well fitted by the linear dependence. Only for optical transitions with a large excitation probability the slow optical pumping process towards other states modifies the linear dependence predicted by the steady state. For some $\sigma/\sigma$ polarization data  theory-experiment discrepancies are observed,  and we attribute them to the presence of not identified CO signals superimposed to the identified ones.\\  
\indent Figure~\ref{Fig5} reports similar comparisons for the CO signals, in (a) for a $\pi/\pi$ spectrum at 0.056 T and in (b) for a $\sigma/\sigma$ spectrum at   0.072 T.  Following our sign convention,  we plot $S_{meas}$ measured for the N signals with a negative sign in order to distinguish them from the VTL signals.  The $f^{VTL}$ theoretical predictions for the VTL positive values are from Eq.~\eqref{VTLamplitude},  and   the $f^{N}$ ones for the N negative values are from Eq.~\eqref{Namplitude}.  The red lines represent theoretical predictions based on the TLS linear dependence reported in Fig.~\ref{Fig4} confirming the correct absolute scale of $S_{meas}$.  Among the  larger number of CO's observed at the lower magnetic field,  several data deviate from the theoretical predictions, and once again we attribute it to the presence of not identified CO's. \\
\begin{figure}[!t]
\centering 
\includegraphics[width=8 cm]{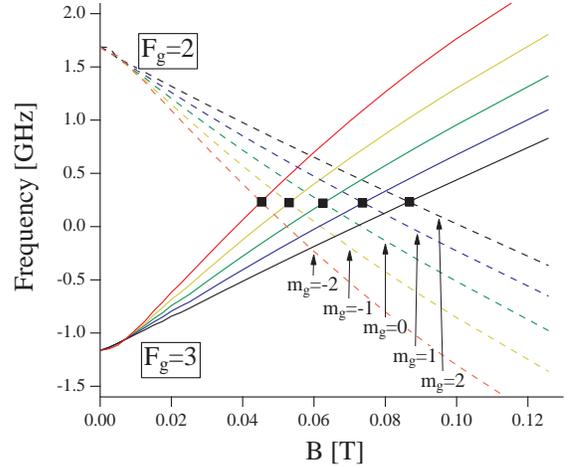}
\caption{Frequencies vs B magnetic field for  $^{85}$Rb optical transition couples producing strong double-N crossovers in $\sigma/\sigma$ driving. Magic field values are denoted by the black squares. Continuous lines for $\sigma^+$ transitions from $|F_g=3,m_g\rangle_g$ states to $|F_e,m_e\rangle_e$ states, and dashed lines for $\sigma^-$ ones  from $|F_g=2,m_g\rangle_g$ ones. In red $|3,2\rangle_g \to |4,3\rangle_e$, $|2,2\rangle_g \to |1,1\rangle_e$; in yellow  $|3,1\rangle_g \to |4,2_e\rangle,  |2,1\rangle_g \to |1,0\rangle_e$; in green  $|3,0\rangle_g \to |4,1\rangle_e$, $|2,0\rangle_g \to |1,-1\rangle_e$, in blue  $|3,-1\rangle_g \to |4,0\rangle_e$, $|2,-1\rangle_g \to |2,-2\rangle_e$; in black  $|3,-2\rangle_g \to |4,-1\rangle_e$, $|2,-2\rangle_g \to |3,-3\rangle_e$. The zero frequency is the centre of gravity of the $B=0$ $^{85}$Rb spectra. At B=0 the transitions  are separated by the 3 GHz ground state hyperfine splitting}. 
\label{Fig6}
\end{figure}
\indent All the N crossover signals with $|f^N|>0.3$ are produced by double-N schemes, and within the  $ \sigma$ spectra intensities larger than the $f_{g-e}=1$ value associated to B=0 cycling transition of $^{87}$Rb are fairly common, see those two present in Fig.~\ref{Fig5}(b). Most CO's of the $\sigma/\sigma$ spectra   are associated to a common level configuration, with ground states within the scheme of Fig.~\ref{Energylevels}(d) as $|1\rangle=|F_g=I-1/2,m_g\rangle$ and $|3\rangle= |F_g=I+1/2,m_g\rangle$, where $I$ is the nuclear spin quantum number, 3/2 and 5/2 for the two Rb isotopes, respectively.  Within the explored magnetic range, those ground states are coupled to excited states by $\sigma^+$/$\sigma^-$ B=0 allowed transition  with $f_{g-e}$ line strengths larger than 0.4. In addition, all the $|2\rangle$ and $|4\rangle$ excited states have branching decay rates very similar, i.e., $\Pi_{21}\approx\Pi_{43}$ between 0.6 and 0.4, and  $\Pi_{23}\approx\Pi_{41}$ between 0.6 and 0.4. Therefore the N signals created by exchanging the action of the control and probe on each N scheme have comparable intensity, contributing to a total strong intensity. Examples of those transitions are listed within the caption of Fig. 6. The $\pi/\pi$ configuration CO's are produced by the Fig.~\ref{Energylevels}(d) scheme with $|1\rangle=|F_g,m_g\rangle$ and  $|3\rangle= |F_g,m_g\pm1\rangle$, the $\Pi_{23}$ and $\Pi_{41}$ decay channels being associated to the $\sigma$ transitions. These CO's have intensities  smaller than those on $\sigma/\sigma$ excitations because the laser driven $\pi$ transitions are weaker than the laser driven $\sigma$ transitions.\\
\indent For all the cases reported in this work the laser driven $|1\rangle \to |2\rangle$ and $|3\rangle \to |4\rangle$ transitions have different frequencies, and the atomic velocity compensates for that difference. For the case of $\omega_c=\omega_p$ the resonant atomic velocity condition of Eq.~\eqref{veloccond2} requires $|\omega_{21}-\omega_{43}| \le 2ku$ i.e. twice the Doppler width of  500 MHz.

\section{Perspective and conclusions}
\indent Our observed double-N crossovers are produced by two different velocity classes as it appears from Eq.~\eqref{veloccond2} imposing $\omega_{21} \ne \omega_{43}$. However by properly tuning the magnetic field it is possible to identify magic values where the condition $\omega_{21}=\omega_{43}$ is satisfied.  Then, for $\omega_p=\omega_c$  at the magic magnetic fields a single atomic zero velocity class interacts simultaneously with the control and probe lasers driving each N scheme of the double-N the signals.  This is shown in Fig.~\ref{Fig6} for the case of $\sigma/\sigma$ driving of the  coupling/probe transitions and for level schemes with $|1\rangle=|F_g=I-1/2,m_g\rangle$ and $|3\rangle= |F_g=I+1/2,m_g\rangle$.   Examples of the frequency dependence on the $B$ magnetic field for those coupled transitions forming the double-N signals are reported there. The magic field values are denoted by black squares. Other combinations with same ground states and different  excited states also produce strong double-N crossovers.  In $^{87}$Rb owing to the larger ground state hyperfine splitting similar double-N schemes are realized at magnetic fields in the  0.1-0.2 T range.  At the magic magnetic field values the absorbing atoms experience a {\it push-pull} evolution between the four levels, similar to the push-pull evolution in double-$\Lambda$ driven systems~\cite{JauHapper2004,Zanon2005}. It will  be interesting to investigate if the transient regime of the absorption signals could present new additional features.  In addition at the magic magnetic field values of Fig.~\ref{Fig6} applying two separate laser radiations driving the  $|1\rangle\to|4\rangle$ and $|3\rangle\to|2\rangle$ transitions a double-$\Lambda$ scheme is realized where two separate atomic transitions are driven by the same laser. For such magic field configuration the processes of electromagnetically induced transparency and slow light  will be enhanced. The magic field values for $^{85}$Rb are within the range explored within the present investigation. It may be difficult to assemble a permanent magnet generating exactly the magic value. Instead that value may be reached more easily by tuning the field with  additional Helmoltz coils.\\ 
\indent We have performed a sub-Doppler spectroscopic analysis of the Rb isotope absorption in magnetic fields within the 0.05-0.13 T range. We have identified a large number of absorbing transitions. In addition the spectra present a large number of crossovers produced by three-level V-scheme and four level N configurations. No $\Lambda$-scheme crossover is detected, but several of them should appear within the mixed $\pi/\sigma$ laser scheme, even if their intensity will be not very large because most transitions are open. Our recorded spectra present very strong saturated  absorption signals produced by double-N configurations where two different atomic velocity classes contribute to the  absorption.\\
\indent The intensities are analyzed through a simple model, providing a complete picture for both two-level and many-level signals. The agreement between theory and experiment may be improved by developing a model of the transient regime for the crossovers and applying the Doppler analysis to both the two-level resonances and  the crossovers.\\  
\indent  In our experimental observations the laser induced population transfers among the levels of N four-level schemes are incoherently produced by spontaneous emission processes.  However, all the explored incoherent N schemes can be transformed into coherent ones by applying an additional resonant laser radiation. Therefore all alkali atoms immersed into medium-high magnetic fields have a large number of strong and easily accessible coherent double-N coherent configurations.  

\section{Acknowledgments}
This research  has been partially supported through the grant NEXT n$^{\rm o}$ ANR-10-LABX-0037 in the framework of the  "Programme des Investissements d'Avenir".  SS acknowledges a financial support from Universit\'e Franco Italienne, and  EA from the Chair d'Excellence Pierre de Fermat of the Conseil Regional Midi-Pyren\'ees. The authors acknowledge the precious technical assistance of Nicola Puccini in realising the magnetic assembly, the preparation by Nicolas Bruyant of the software recording simultaneously data from absorption, FP and wavemeter, and the support by Remy Battesti and Mathilde Fouch\'e. EA is very grateful to Renaud Mathevet for sharing his atomic physics knowledge.

   \end{document}